\documentstyle[12pt]{article}
\input epsf

\global\arraycolsep=2pt

\begin{document}

\begin{titlepage}

\begin{flushright}
CERN-TH.7396/94\\
hep-ph/9409306
\end{flushright}

\vspace{0.3cm}

\begin{center}
\Large\bf Theoretical Uncertainties in the\\
Extraction of {\boldmath $|V_{cb}|$}\\
from {\boldmath $\bar B\to D^*\ell\,\bar\nu$} Decays near Zero Recoil
\end{center}

\vspace{0.8cm}

\begin{center}
Matthias Neubert\\
{\sl Theory Division, CERN, CH-1211 Geneva 23, Switzerland\\
(E-mail: neubert@cernvm.cern.ch)}
\end{center}

\vspace{0.8cm}

\begin{abstract}
I discuss the theoretical uncertainties in the extraction of
$|\,V_{cb}|$ from a measurement of the $\bar B\to D^*\ell\,\bar\nu$
decay rate close to zero recoil. In particular, I combine previous
estimates of the $1/m_Q^2$ corrections to the normalization of the
hadronic form factor at zero recoil with sum rules derived by Shifman
{\it et al}.\ to obtain a new prediction with less uncertainty. I
also give a prediction for the slope of the form factor
$\widehat\xi(w)$ at zero recoil: $\widehat\varrho^2=0.7\pm 0.2$.
Using the most recent experimental results, I obtain the
model-independent value $|\,V_{cb}|=0.0395\pm 0.0030$.
\end{abstract}

\bigskip\bigskip

\begin{center}
\it Invited talk presented at the 27th International Conference on
High Energy Physics\\
Glasgow, Scotland, 20--27 July 1994
\end{center}

\bigskip\bigskip

\noindent
CERN-TH.7396/94\\
September 1994

\end{titlepage}

\section{Introduction}

With the discovery of heavy quark symmetry (for a review see
Ref.~\cite{review} and references therein), it has become clear that
the study of exclusive semileptonic $\bar B\to D^*\ell\,\bar\nu$
decays close to zero recoil allows for a reliable determination of
the Cabibbo--Kobayashi--Maskawa matrix element $V_{cb}$, which is
free, to a large extent, of hadronic uncertainties
\cite{Volo}--\cite{Vcb}. Model dependence enters this analysis only
at the level of power corrections, which are suppressed by a factor
of at least $(\Lambda_{\rm QCD}/m_c)^2$. These corrections can be
investigated in a systematic way using the heavy quark effective
theory \cite{Geor}. They are found to be small, of the order of a few
per cent.

Until recently, this method to determine $|\,V_{cb}|$ was limited by
large experimental uncertainties of about 15--20\%, which were much
larger than the theoretical uncertainties in the analysis of
symmetry-breaking corrections. However, three collaborations have now
presented results of higher precision \cite{CLEO}--\cite{ARGUS}. It
is thus important to reconsider the status of the theoretical
analysis, even more so since the original calculation of power
corrections in Ref.~\cite{FaNe} has become the subject of controversy
\cite{Shif}.

The differential decay rate for the process $\bar B\to D^*\ell\,
\bar\nu$ is given by \cite{review}
\begin{eqnarray}\label{BDrate}
   {{\rm d}\Gamma\over{\rm d}w}
   &=& {G_F^2\over 48\pi^3}\,(m_B-m_{D^*})^2\,m_{D^*}^3
    \sqrt{w^2-1}\,(w+1)^2 \nonumber\\
   &&\mbox{}\times \bigg[ 1 + {4w\over w+1}\,
    {m_B^2-2w\,m_B m_{D^*} + m_{D^*}^2\over(m_B - m_{D^*})^2}
    \bigg]\,|\,V_{cb}|^2\,\eta_A^2\,\widehat\xi^{\,2}(w) \,,
\end{eqnarray}
where
\begin{equation}
   w = v_B\cdot v_{D^*} = {m_B^2 + m_{D^*}^2 - q^2
   \over 2 m_B m_{D^*}}
\end{equation}
denotes the product of the meson velocities. I have factorized the
hadronic form factor for this decay into a short-distance coefficient
$\eta_A$ and a function $\widehat\xi(w)$, which contains the
long-distance hadronic dynamics. Apart from corrections of order
$1/m_Q$, this function coincides with the Isgur--Wise form factor
\cite{Isgu,Falk}. Luke's theorem determines the normalization of
$\widehat\xi(w)$ at zero recoil ($w=1$) up to second-order power
corrections \cite{Vcb,Luke}:
\begin{equation}
   \widehat\xi(1) = 1 + \delta_{1/m^2} \,.
\end{equation}
The strategy is to obtain the product $|\,V_{cb}|\,\eta_A\,
\widehat\xi(w)$ from a measurement of the differential decay rate,
and to extrapolate it to $w=1$ to extract
\begin{equation}
   |\,V_{cb}|\,\eta_A\,(1 +  \delta_{1/m^2})
   = |\,V_{cb}|\,\Big\{ 1 + O(\alpha_s,1/m_Q^2) \Big\} \,.
\end{equation}
The task of theorists is to provide a reliable calculation of
$\eta_A$ and $\delta_{1/m^2}$ in order to turn this measurement into
a precise determination of $|\,V_{cb}|$.

\section{Calculation of $\eta_A$}

The short-distance coefficient $\eta_A$ takes into account a finite
renormalization of the axial vector current in the region
$m_b>\mu>m_c$. Its calculation is a straightforward application of
QCD perturbation theory. At the one-loop order, one finds
\cite{Volo,QCD1}
\begin{equation}
   \eta_A = 1 + {\alpha_s\over\pi}\,\bigg(
   {m_b + m_c\over m_b - m_c}\,\ln{m_b\over m_c} - {8\over 3}
   \bigg) \,.
\end{equation}
The scale of the running coupling constant is not determined at this
order. Choosing $\alpha_s$ between $\alpha_s(m_b)\simeq 0.20$ and
$\alpha_s(m_c)\simeq 0.32$, and using $m_c/m_b=0.30\pm 0.05$, one
obtains values in the range $0.95<\eta_A<0.98$. The scale ambiguity
leads to an uncertainty of order $\Delta\eta_A\sim[(\alpha_s/\pi)
\ln(m_b/m_c)]^2\sim 2\%$.

The calculation can be improved by using the renormalization group to
resum the leading and next-to-leading logarithms of the type
$[\alpha_s\ln(m_b/m_c)]^n$, $\alpha_s [\alpha_s\ln(m_b/m_c)]^n$, and
$(m_c/m_b) [\alpha_s\ln(m_b/m_c)]^n$ to all orders in perturbation
theory \cite{PoWi}--\cite{FaGr}. A consistent scheme for a
next-to-leading-order calculation of $\eta_{\em A}$ has been
developed in Ref.~\cite{QCD2}. The result is
\begin{eqnarray}
   \eta_A = x^{6/25}\,\bigg\{ 1
   &+& 1.561\,{\alpha_s(m_c)-\alpha_s(m_b)\over\pi}
    - {8\alpha_s(m_c)\over 3\pi} \nonumber\\
   &+& {m_c\over m_b}\,\bigg( {25\over 54}
    - {14\over 27}\,x^{-9/25} + {1\over 18}\,x^{-12/25}
    + {8\over 25}\,\ln x \bigg) \nonumber\\
   &+& {2\alpha_s(m)\over\pi}\,
    {m_c^2\over m_b(m_b - m_c)}\,\ln{m_b\over m_c} \bigg\} \,,
\end{eqnarray}
where $x=\alpha_s(m_c)/\alpha_s(m_b)$, and $m_b>m>m_c$. The numerical
result is very stable under changes of the input parameters. For
$\Lambda_{\overline{\rm MS}}=(0.25\pm 0.05)$ GeV (for 4 flavours) and
$m_c/m_b=0.30\pm 0.05$, one obtains $\eta_A=0.985\pm 0.006$. The
uncertainty arising from next-to-next-to-leading corrections is of
order $\Delta\eta_A\sim(\alpha_s/\pi)^2\sim 1\%$. Taking this into
account, I think it is conservative to increase the error by a factor
2.5 and quote
\begin{equation}
   \eta_A = 0.985\pm 0.015 \,.
\end{equation}

\section{Anatomy of $\delta_{1/m^2}$}

Hadronic uncertainties enter the determination of $|\,V_{cb}|$ at
the level of second-order power corrections, which are expected to be
of order $(\Lambda_{\rm QCD}/m_c)^2\sim 3\%$. For a precision
measurement, it is important to understand the structure of these
corrections in detail. Falk and myself have derived the exact
expression \cite{FaNe}
\begin{equation}\label{delm2}
   \delta_{1/m^2} = - \bigg( {1\over 2 m_c} - {1\over 2 m_b}
   \bigg) \bigg( {\ell_V\over 2 m_c} - {\ell_P\over 2 m_b}
   \bigg) + {1\over 4 m_c m_b}\,\bigg( {4\over 3}\,\lambda_1
   + 2\lambda_2 - \lambda_{G^2} \bigg) \,,
\end{equation}
which depends upon five hadronic parameters: $\ell_P$ and $\ell_V$
parametrize the deficit in the ``wave-function overlap'' between $b$-
and $c$-flavoured pseudoscalar ({\it P\/}) and vector ({\it V\/})
mesons, for instance
\begin{equation}
   \langle D(v)|\,c^\dagger b\,|B(v)\rangle\propto 1 -
   \bigg( {1\over 2 m_c} - {1\over 2 m_b} \bigg)^2\,\ell_P \,.
\end{equation}
The parameter $-\lambda_1=\langle\vec p_Q^{\,2}\rangle$ is
proportional to the kinetic energy of the heavy quark inside a heavy
meson, $\lambda_2=(m_{B^*}^2 - m_B^2)/4\simeq 0.12$ GeV$^2$ is
determined by the vector--pseudoscalar mass splitting, and
$\lambda_{G^2}$ parametrizes certain matrix elements containing
double insertions of the chromo-magnetic operator. With the exception
of $\lambda_2$, estimates of these parameters are model-dependent. In
Ref.~\cite{FaNe}, we made the simplifying assumptions that
$\ell_P=\ell_V$, and that the corrections represented by
$\lambda_{G^2}$ are negligible. The latter one is based on the
observation that these corrections involve a double insertion of an
operator that breaks the heavy quark spin symmetry. Using then
reasonable values such as $\ell_P=\ell_V=(0.35\pm 0.15)$ GeV$^2$ and
$-\lambda_1=(0.25\pm 0.20)$~GeV$^2$, one obtains
$\delta_{1/m^2}=-(2.4\pm 1.3)\%$. Here and in the following, I take
$m_b=4.80$ GeV and $m_c=1.45$ GeV for the heavy quark masses. In
Ref.~\cite{review}, the error in the estimate of $\delta_{1/m^2}$ has
been increased to $\pm 4\%$ in order to account for the model
dependence and higher-order corrections. A very similar result,
$-5\%<\delta_{1/m^2}<0$, has been obtained by Mannel \cite{Mann}.

Recently, Shifman {\it et al}.\ have suggested an alternative
approach to obtain an estimate of $\delta_{1/m^2}$, which is based on
bounds derived using sum rules and the operator product expansion
\cite{Shif}. These bounds imply the inequalities
\begin{eqnarray}\label{sumruls}
   \ell_P &>& {1\over 2}(-\lambda_1 - 3\lambda_2)
    \equiv \ell_P^{\rm min} > 0 \,, \nonumber\\
   \ell_V &>& {1\over 2}(-\lambda_1 + \lambda_2)
    \equiv \ell_V^{\rm min} > 2\lambda_2 \,, \nonumber\\
   \delta_{1/m^2} &<& - \bigg( {1\over 2 m_c} - {1\over 2 m_b}
    \bigg) \bigg( {\ell_V^{\rm min}\over 2 m_c}
    - {\ell_P^{\rm min}\over 2 m_b} \bigg)
    + {1\over 4 m_c m_b}\,\bigg( {4\over 3}\,\lambda_1
    + 2\lambda_2 \bigg) \nonumber\\
   &<& - {\lambda_2\over 2 m_c^2}\simeq -2.9\% \,.
\end{eqnarray}
The upper bound for $\delta_{1/m^2}$ implies that $\eta_A\,
\widehat\xi(1) < 0.956$. In Ref.~\cite{Shif}, this number is quoted
as 0.94. In the same reference, the authors give an ``educated
guess'' $\eta_A\,\widehat\xi(1) = 0.89\pm 0.03$ corresponding to
$\delta_{1/m^2}=-(9.6\pm 3.0)\%$. However, the arguments presented to
support this guess are not very rigorous.

It is possible to combine the above approaches to reduce the
theoretical uncertainty in the estimate of $\delta_{1/m^2}$
\cite{new}. The idea is to use the sum rules to constrain the
hadronic parameters in (\ref{delm2}) in a threefold way: (i)~The
first relation in (\ref{sumruls}) implies that
\begin{equation}\label{lam1bou}
   -\lambda_1 > 3\lambda_2\simeq 0.36~\mbox{GeV}^2 \,,
\end{equation}
excluding some of the values for the parameter $\lambda_1$ used in
previous analyses. (ii)~Comparing the third relation in
(\ref{sumruls}) with (\ref{delm2}) in the limit $m_b=m_c$, one finds
that
\begin{equation}
   \lambda_{G^2} > 0 \,.
\end{equation}
(iii) Finally, $\ell_P$ and $\ell_V$ are correlated in such a way
that $\ell_V>\ell_P$ if $\lambda_{G^2}$ is not too large. To
illustrate this last point, let me define new parameters
\begin{equation}
   \bar\ell = {1\over 2}\,(\ell_V + \ell_P) \,,\qquad
   D = {1\over 2}\,(\ell_V - \ell_P) - \lambda_2 \,.
\end{equation}
In terms of these,
\begin{eqnarray}\label{final}
   \delta_{1/m^2} &=& - \bigg( {1\over 2 m_c} - {1\over 2 m_b}
    \bigg)^2\,\bar\ell - \bigg( {1\over 4 m_c^2} - {1\over 4 m_b^2}
    \bigg) (\lambda_2 + D) \nonumber\\
   &&\mbox{}+ {1\over 4 m_c m_b}\,\bigg( {4\over 3}\,\lambda_1
    + 2\lambda_2 - \lambda_{G^2} \bigg) \,.
\end{eqnarray}
Using the inequalities (\ref{sumruls}), one can show that
$\bar\ell>{1\over 2} (-\lambda_1-\lambda_2)$ and
$-D_{\rm max}<D<D_{\rm max}$, with \cite{new}
\begin{equation}
   D_{\rm max} = \left\{ \begin{array}{cl}
    S & ;\quad 0<S\le\lambda_{G^2}/2 \,, \\
    \sqrt{\lambda_{G^2} S - \lambda_{G^2}^2/4} &
    ;\quad S\ge\lambda_{G^2}/2 \,,
   \end{array} \right.
\end{equation}
where $S=\bar\ell + (\lambda_1+\lambda_2)/2$. There are thus three
effects, which decrease $\delta_{1/m^2}$ with respect to the estimate
given in Ref.~\cite{FaNe}: a large value of $(-\lambda_1)$, a
positive value of $\lambda_{G^2}$, and the fact that for small
$\lambda_{G^2}$ the difference $(\ell_V-\ell_P)$ is centred around
$2\lambda_2$ (i.e.\ $D$ is centred around 0).

\begin{figure}[t]
   \epsfxsize=9.5cm
   \centerline{\epsffile{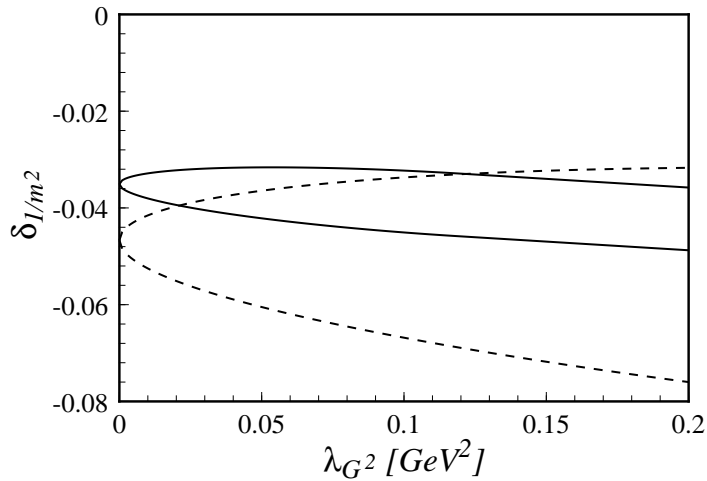}}
{\caption{\label{fig1} Allowed regions for $\delta_{1/m^2}$ as a
function of $\lambda_{G^2}$ for the two cases $\bar\ell=0.2$ GeV$^2$
(solid) and 0.4 GeV$^2$ (dashed).}}
\end{figure}

In evaluating (\ref{final}), I will take $-\lambda_1=0.4$ GeV$^2$,
which is consistent with the bound in (\ref{lam1bou}) and with the
value $-\lambda_1=(0.5\pm 0.1)$ GeV$^2$ obtained from QCD sum rules
\cite{BaBr}. Varying $-\lambda_1$ in the range between 0.36 and 0.5
GeV$^2$ does not alter the results very much. The main uncertainty
comes from the unknown values of the parameters $\bar\ell$ and
$\lambda_{G^2}$. As a guideline, one may employ the constituent quark
model of Isgur {\it et al}.\ \cite{ISGW}, in which one uses
non-relativistic harmonic oscillator wave functions for the
ground-state heavy mesons, for instance $\psi_B(r)\sim \exp(-{1\over
2}\mu\omega r^2)$, where $\mu=(1/m_q+1/m_b)^{-1}$ is the reduced
mass. One then obtains $\bar\ell={3\over 4} m_q^2\simeq 0.2$ GeV$^2$,
where I use $m_q\simeq 0.5$ GeV for the light constituent quark mass.
However, this estimate of $\bar\ell$ is probably somewhat too low.
Lattice studies of heavy-light wave functions suggest an exponential
behaviour of the form $\psi_B(r)\sim\exp(-\kappa\mu r)$ \cite{latt},
which leads to $\bar\ell={3\over 2} m_q^2\simeq 0.4$ GeV$^2$. Values
much larger than this are unlikely, since I use a rather large
constituent quark mass $m_q$. In fact, adopting the point of view
that the sum rules for $\ell_P$ and $\ell_V$ are saturated to
approximately 50\% by the ground-state contribution \cite{Shif}, one
would expect $\bar\ell\simeq (-\lambda_1-\lambda_2)\simeq 0.28$
GeV$^2$, which seems a very reasonable value to me. In
Fig.~\ref{fig1}, I show the allowed regions for $\delta_{1/m^2}$ as a
function of $\lambda_{G^2}$ for two values of $\bar\ell$. I think it
is reasonable to assume that $\lambda_{G^2}$ is of a magnitude
similar to $\lambda_2$ or smaller. Thus, I conclude that for all
reasonable choices of parameters the results are in the range
\begin{equation}
   -8\% < \delta_{1/m^2} < -3\% \,,
\end{equation}
which is consistent with the previous estimates in
Refs.~\cite{FaNe,Shif,Mann} at the $1\sigma$ level. A more precise
deter\-mination of the parameter $\bar\ell$ would help to reduce the
uncertainty in this number.

\section{Prediction for the slope parameter $\widehat\varrho^2$}

In the extrapolation of the differential decay rate (\ref{BDrate})
to zero recoil, the slope of the function $\widehat\xi(w)$ close to
$w=1$ plays an important role. One defines a parameter
$\widehat\varrho^2$ by
\begin{equation}
   \widehat\xi(w) = \widehat\xi(1)\,\Big\{ 1
   - \widehat\varrho^2\,(w-1) + \ldots \Big\} \,.
\end{equation}
It is important to distinguish $\widehat\varrho^2$ from the slope
parameter $\varrho^2$ of the Isgur--Wise function. They differ by
corrections that break the heavy quark symmetry. Whereas the slope of
the Isgur--Wise function is a universal, mass-independent parameter,
the slope of the physical form factor depends on logarithms and
inverse powers of the heavy quark masses. On the other hand,
$\widehat\varrho^2$ is an observable quantity, while the value of
$\varrho^2$ depends on the renormalization scheme. To illustrate this
last point, let me neglect for the moment $1/m_Q$ corrections and
work in the leading logarithmic approximation. Then the relation
between the physical slope parameter $\widehat\varrho^2$ and the
slope parameter $\varrho^2(\mu)$ of the regularized Isgur--Wise
function is \cite{new}
\begin{equation}
   \widehat\varrho^2 = \varrho^2(\mu) - {16\over 81}\,
   \ln {\alpha_s(m)\over\alpha_s(\mu)} \equiv
   \varrho^2 - {16\over 81}\,\ln\alpha_s(m) \,,
\end{equation}
where $\mu$ is the renormalization scale, and $m$ is an undetermined
(at this order) scale between $m_b$ and $m_c$. The last equation can
be used to define the $\mu$-independent slope of the renormalized
Isgur--Wise function (see Ref.~\cite{QCD2} for the generalization of
this definition to next-to-leading order). If next-to-leading
logarithmic corrections are taken into account, the scale ambiguity
related to the choice of $m$ is resolved, and one obtains
\begin{equation}\label{rhorel}
   \widehat\varrho^2 = \varrho^2 + (0.14\pm 0.02) + O(1/m_Q) \,.
\end{equation}
The $1/m_Q$ corrections to this relation have been investigated and
are found to be negative. However, any such estimate is
model-dependent and thus has a large theoretical uncertainty. The
result is $\widehat\varrho^2\simeq\varrho^2\pm 0.2$ \cite{new}.
Predictions for the renormalized slope parameter $\varrho^2$ are
available from QCD sum rules including a next-to-leading-order
renormalization-group improvement. One obtains $\varrho^2\simeq
0.7\pm 0.1$ \cite{review,Baga,twoloop}. I thus expect
\begin{equation}
   \widehat\varrho^2 = 0.7\pm 0.2 \,.
\end{equation}

\section{Summary}

Using the updated values $\eta_A=0.985\pm 0.015$ and $\delta_{1/m^2}
= -(5.5\pm 2.5)\%$, I obtain for the normalization of the hadronic
form factor at zero recoil:
\begin{equation}\label{etaxi}
   \eta_A\,\widehat\xi(1) = 0.93\pm 0.03 \,.
\end{equation}
Three experiments have recently presented new measurements of the
product $|\,V_{cb}|\,\eta_A\,\widehat\xi(1)$. When rescaled using the
new lifetime values $\tau_{B^0}=(1.61\pm 0.08)$~ps and
$\tau_{B^+}=(1.65\pm 0.07)$ ps \cite{Roud}, the results obtained from
a linear fit to the data are
\begin{equation}
   |\,V_{cb}|\,\eta_A\,\widehat\xi(1) = \left\{
   \begin{array}{ll}
   0.0347\pm 0.0019\pm 0.0020 &
    ;\quad \mbox{Ref.~\protect\cite{CLEO},} \\
   0.0364\pm 0.0042\pm 0.0031 &
    ;\quad \mbox{Ref.~\protect\cite{ALEPH},} \\
   0.0385\pm 0.0043\pm 0.0028 &
    ;\quad \mbox{Ref.~\protect\cite{ARGUS},}
   \end{array} \right.
\end{equation}
where the first error is statistical and the second systematic. I
will follow the suggestion of Ref.~\cite{Ritch} and add $0.001\pm
0.001$ to these values to account for the curvature of the function
$\widehat\xi(w)$. Taking the weighted average of the experimental
results and using the theoretical prediction (\ref{etaxi}), I then
obtain
\begin{equation}
   |\,V_{cb}| = 0.0395\pm 0.0027\,(\mbox{exp})
   \pm 0.0013\,(\mbox{th}) = 0.0395\pm 0.0030 \,,
\end{equation}
which corresponds to a model-independent measurement of $|\,V_{cb}|$
with 7\% accuracy. This is by far the most accurate determination to
date.

Neglecting $1/m_Q$ corrections, I have related the physical slope
parameter $\widehat\varrho^2$ to the slope of the Isgur--Wise
function and obtain the prediction $\widehat\varrho^2=0.7\pm 0.2$. It
compares well with the average value observed by experiments, which
is $\widehat\varrho^2=0.87\pm 0.12$ \cite{CLEO}--\cite{ARGUS}.

\end{document}